\documentclass{aa}
\usepackage{txfonts}
\usepackage{graphicx}
\begin{document}
   \title{The Globular Cluster System of \object{NGC\,1399} IV. Some noteworthy objects}

   \author{T. Richtler
          \inst{1}
          \and
          B. Dirsch
          \inst{1}
          \and
          S. Larsen
          \inst{2}
          \and
          M. Hilker
          \inst{3}
          \and
          L. Infante
          \inst{4}
      \thanks{
       Based on observations collected with the VLT at the European Southern 
        Observatory, Cerro Paranal, Chile; ESO program  66.B-0393}
           }

   \offprints{T. Richtler}

   \institute{
             Grupo de Astronom\'{\i}a, Departamento de F\'{\i}sica, 
                Universidad de Concepci\'on, Casilla 160-C,
                Concepci\'on, Chile\\
             \email{tom@coma.cfm.udec.cl}\\
             \and
             European Southern Observatory,
             Karl-Schwarzschild-Str.2,
             D-85748 Garching, Germany\\
             \and
             Sternwarte der Universit\"at Bonn,
             Auf dem H\"ugel 71, D-53121\\
             \and
             Departamento de Astronom\'{\i}a y Astrof\'{\i}sica, P. Universidad Cat\'olica,
             Vicu\~na Mackenna 4860, Santiago 22, Chile\\
             }

   \date{Received / Accepted}

   \abstract{We present 8 bright globular clusters and/or objects of less familiar nature
which we found in the course of scrutinizing the globular cluster system of NGC 1399.
These objects are morphologically striking, either by their sizes or by other structural
properties. Some of them may be candidates for stripped dwarf galaxy nuclei, emphasizing
the possible role of accretion in the NGC 1399 cluster system. 
They are all highly interesting targets for
further deep spectroscopy  or HST-imaging. Since these objects have been found within an
area of only 42$\arcmin^2$, we expect many more still to be detected in a full census
of the NGC 1399 cluster system.

}

   \maketitle
%
%________________________________________________________________

\section{Introduction}

NGC 1399 is the central galaxy in the
Fornax cluster.  
Due to its proximity (19 Mpc)
and richness, its globular cluster system is one of the most attractive cluster systems  to study
(for a compilation of literature on NGC 1399 see the introduction of
Dirsch et al. \cite{dirsch2003}). 

How such rich GCSs of giant elliptical galaxies have formed, is intimately
linked to the formation history of the host galaxy itself. Several possibilities
have been discussed: Tidal stripping of globular clusters (GCs) by encounters
of neighbouring galaxies (Kissler-Patig et al. \cite{kissler1999}), accretion of GCs 
through the accretion of dwarf galaxies (Hilker et al. \cite{hilker1999a}, 
C\^ot\'e et al. \cite{cote1998}), formation of
GCs in merger events (Ashman \& Zepf \cite{ashman92}), in-situ formation during collapse
(Forbes et al. \cite{forbes97}).

In particular the brightest objects attract closer scrutiny. In NGC 1399, 
GC-like objects as bright as $M_V \approx -13$ mag have been identified 
(Hilker et al. \cite{hilker1999}, Drinkwater et al. \cite{drink2000}) which
Phillips et al. (\cite{phillips01}) labelled ''Ultracompact Dwarfs'' (UCDs). Do these objects simply constitute 
the bright wing of the globular cluster luminosity function (Mieske et
al. \cite{mieske2002}) or 
were they formed by processes different from those forming ''normal''
globular clusters? 
Recent spectroscopic identification of new compact Fornax members around NGC 1399 by Mieske et al.
 (\cite{mieske2004}) rather support the latter hypothesis by finding an overdensity for bright
objects with respect to the globular cluster luminosity function.  
Accretion of dwarf galaxies as donators of globular clusters has first been considered by
Zinnecker et al. (\cite{zinnecker1988}). Meanwhile, much evidence for this has been found
in the Galactic system with the cases of the Sagittarius dwarf galaxy (e.g. Ibata et al. 
\cite{ibata1997})  and $\omega$ Centauri
(e.g. Hilker et al. \cite{hilker2004} and references therein). Some of the brightest clusters in NGC 5128
are also suspected to be stripped nuclei of dwarf galaxies (Martini \& Ho \cite{martini2004}).

If nuclei of 
former dwarf galaxies make up a significant proportion of the bright
objects in NGC 1399 (Karick et al. \cite{karick2003}, Mieske et al. \cite{mieske2004}), one  expects to find evidence for that for example in form of
faint appendages hinting to tidal processes.

We studied the GCS of NGC 1399 both photometrically, by
wide-field imaging (Dirsch et al. \cite{dirsch2003})
and spectroscopically (Richtler et al. \cite{richtler2004}, Dirsch et al. \cite{dirsch2004}) 
to a larger extent as has been done before. In the course of this campaign, we have gained  
VLT images in excellent seeing, resolving many globular clusters and revealing structures
 which went unnoticed until now. 
 
In this paper, we present a list of some of our brightest GCs showing under
various aspects some sort of peculiarity regarding their sizes or other structural properties.
Our aim is to draw the
community's attention to these interesting objects in order to investigate
their properties further by  better spatial or spectroscopic resolution.  
%Since these objects have been found within the  area of two FORS-fields, i.e. only 42$\arcmin^2$,
%a complete census of the cluster system of NGC 1399 is expected to reveal a multitud of interesting
%objects.

\section{Observations}

The spectroscopic observations, which resulted in radial velocities of about 470 objects,
 have been obtained with FORS2 and the
Mask Exchange Unit (MXU) at ESO's Very Large Telescope in the period 11/29/2000-
12/1/2000. 
The grism was 600B, resulting in a resolution of about 4 \AA. These
observations are described in Dirsch et al. \cite{dirsch2004}. During this
campaign, two FORS2 fields (area 6.5$\times$6.5 $\arcmin^2$) have been imaged in V and I  in excellent
seeing (0.6 arcsec), resulting in marginal resolution of many GCs. 
These images are already documented  in Dirsch et al. \cite{dirsch2003}. The images
have been exposed for 300 sec in both V (Bessell V) and  I (Bessell I).
 A few objects
are well resolved and in some cases show faint large extensions or irregular
shapes. These are the objects, which we list here. The quoted Washington
photometric values are taken from Dirsch et al.  \cite{dirsch2003}, who used
the 4m CTIO telescope equipped with the MOSAIC camera to obtain wide-field photometry
in Washington C1 and Kron-Cousins R.

\section{Morphological description}
The objects which we present have been  found by visual inspection of the
 PSF-subtracted
images. Most faint structure are better or only visible on the V-frame due to the much lower
sky background with respect to the I frame. The PSF subtraction works excellent, removing also
bright stars and leaving only faint residuals. Many of the bright  globular clusters, which
can be spectroscopically identified, leave much brighter residuals than stars of comparable or
even distinctly higher brightness. 
Of course, the resolution is not good enough to measure except  for the
largest ones reliable structural parameters, but it demonstrates that large GCs can be identified by
ground-based data out to a distance of 20 Mpc.  
In NGC 1399, Larsen et al. (\cite{larsen2001}) measured effective radii with HST, but
the present FORS fields do not overlap with the HST-fields. 
Here we select only objects which are particularly striking. 

For estimating Washington metallicities, we used the calibration formula given by Harris \& Harris (\cite{harris2002}):
$$  [Fe/H] = -6.04\cdot (1-0.82(C-T1)+0.16(C-T1)^2). $$

\begin{table*}
\caption{The identification of objects uses the numbering of Dirsch et al. (\cite{dirsch2004}). The
label FCOS is adapted from Mieske et al. (\cite{mieske2002},\cite{mieske2004}), the label UCD from
Drinkwater et al. (\cite{drink2003}). We list coordinates and Washington colours
 as well from Dirsch et al.. They are not available for two objects. V and I photometry may not be more accurate
 than 0.1 mag, since no nightly calibration
has been done. We used the standard zeropoints and neglected colour terms. Three  objects are found only
on the preimages and have no V,I photometry assigned. Radial velocities are from Dirsch et al..
Remarks refer to crossidentifications or peculiarities.}
\label{objects.tab}
\begin{tabular}{ccccccccc}
\hline
\hline
object  & RA[2000] & Dec[2000] & T1 & C-T1 & V & V-I & rad.vel.& Remarks \\
\hline
75:85 & 3 38 50.37 & -35 22 07.8 & 20.83$\pm$0.02 & 1.78$\pm$ 0.03 & 21.45  &1.14  & 1609$\pm$24 & extended\\

78:12 & 3 38 58.49 & -35 26 28.1 & 19.92$\pm$0.02 & 1.26$\pm$0.03 & 20.54 & 0.84& 1048$\pm$18 & resolved, appendix\\

80:115 & 3 38 16.64 & -35 20 22.9 & 19.78$\pm$0.03 & 1.40$\pm$0.04 &  & & 1432$\pm$20 & only on V-preimage, large\\

%84:64 &  3 38 54.7 & -35 30 16.6 & 21.02$\pm$0.02 & 1.35$\pm$0.03 & & & 1067$\pm$37 & only on V-preimage \\

%86:84 &  3 38 39.53 & -35 29 16.6 & 20.70$\pm$0.02 & 1.38$\pm$0.03 & & & 1088$\pm$22 & only on V-preimage\\

89:22 & 3 38 17.53 & -35 33 02.5 & ? & ? & 20.49 & 0.85  & 1514 $\pm$26 & well resolved\\ 

89:33 & 3 38 19.02 & -35 32 22.1 & 20.52$\pm$0.02 & 1.60$\pm$0.03 &21.2 & 0.97 & 1650$\pm$34 & \\

90:12 & 3 38 14.80 & -35 33 39.5 & ? & ? & 21.18 & 0.79 & 1565$\pm$24 & large!, FCOS 2-2072 \\

%91:41 & 3 37 56.84 & -35 31 51.0 & 20.84$\pm$0.03 & 1.20$\pm$0.03& 21.28 & 0.82& 1484$\pm$19& well resolved \\

92:74 & 3 38 00.16 & -35 30 08.5 & 20.57$\pm$0.04 & 1.46$\pm$0.04& 21.09 & 0.93&  820$\pm$10 & on galaxy?, FCOS 2-2100 \\

91:93 & 3 38 06.27 & -35 28 58.7 &  18.80$\pm$0.04& 1.62$\pm$0.05 & 19.57 & 0.93 & 1239$\pm$14& UCD2, FCOS 2-2111 \\

FCOS 1-063 & 3 38 56.14 & -35 24 49.1 & 19.79$\pm$0.02 & 1.37$\pm$0.03& 20.54 &0.94  & 688$\pm$45 & \\
\hline

\end{tabular}
\end{table*}

\begin{figure}
\includegraphics[width=6cm]{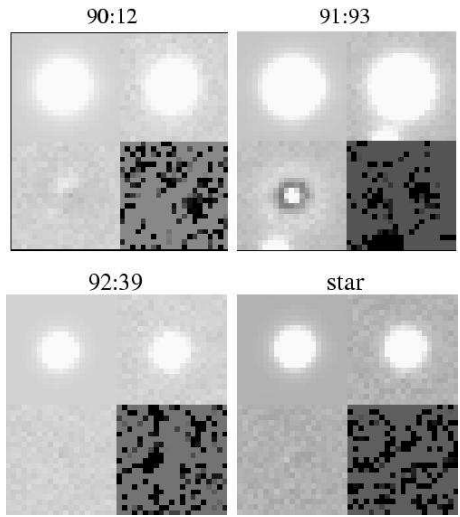}
\caption{The figure shows the ISHAPE-residuals for 90:12 and 91:93 (upper panels). Displayed in 
each panel are the model convolved
by the PSF (upper left), the object (upper right), the residual (lower left), and the weighting
array (lower right). The residuals show that 90:12 is somewhat asymmetric and that a King30 profile
might be not the best representation of 91:93. For comparison, we also show the residuals for 
an unresolved globular cluster (92:39, lower left) and a star of approximately R=19 mag. 
}
\label{ishape}
\end{figure}

\subsection{ISHAPE}
ISHAPE is a code designed to measure the best fitting shape parameters of an object
by iteratively convolving an analytic model with the point spread function (PSF)
(Larsen \cite{larsen1999}). Here we choose a King model with a concentration parameter
of 30, which has been applied to HST-observations of NGC 1399 clusters before (Larsen et al.
 \cite{larsen2001}) (named King30-profile in the following). The model reads
$$ \mu (r) \sim \big(\frac{1}{\sqrt{1+(\frac{r}{r_c})^2}} - \frac{1}{\sqrt{1+(\frac{r_t}{r_c})^2}}\big)^2, $$
$\mu(r)$ being the surface brightness, $r_c$ the core radius, and $r_t$ the tidal radius.

 The code gives the full-width-at-half-maximum (FWHM) and chi-square values of
the model fit. Moreover,  it returns the  chi-square values of
a convolution of a delta-function with the PSF.  This is
useful for an assessment, how well a given object is resolved.
 In our case the FWHM
is always much smaller than the PSF (except for one object), so the resulting values cannot be accurate, but
rather first guesses of their magnitudes. For all objects, we used a fitting radius of 10 pixels.
Table \ref{ishape.tab} lists the values for our objects. The effective radii are of course model dependent and
have been calculated by adopting $r_t$ as the radius containing the total brightness. 
The corresponding central surface brightnesses are very uncertain and only can indicate a trend.
More comments are given in the individual
descriptions. For comparison, we give the fit values also for an unresolved globular cluster
 (92:39)
and a 19th mag (R-band) star. In both cases ISHAPE returns a stellar appearance. Here the chi-square values
of the King30 models are meaningless.  The delta-function chi-square values are very small, indicating an
excellent fit.  
Figure \ref{ishape} illustrates for 90:12 and 91:93 (UCD2), how the quality of the fits can be assessed. 
The upper left panel for each object displays the model, the upper right panel the image, the lower left
panel the residual, and the lower right panel the weighting scheme (black means low weight). More comments
are given in the individual descriptions. In addition, we give the residuals for the two unresolved objects. 
Only tiny residuals are left.

% For 90:12, it
%can be seen that the spherically symmetric model leaves a residual. Apparently, the object is asymmetric. 
%For 91:93, the residual is symmetric but a King30 model underestimates the brightness of the central parts.

\begin{table*}
\caption{This table lists the results of ISHAPE and the corresponding effective radii and central 
surface brighnesses in the V-band. The latter values are rough estimates only, using
the relations quoted by Larsen (\cite{larsen01}). A King30 model has been
applied. The difference between the $\chi^2$-values for a King profile and a Delta-function indicates the degree
of resolution. 
}
\label{ishape.tab}
\begin{tabular}{ccccc}
\hline
\hline
object  & $\chi^2$(King-model/Delta-function) & FWHM [$\arcsec$] & $r_{eff}$[pc] & $\mu_0(V)[mag/arcsec^2]$ \\
\hline
75:85 &  marginally resolved   &  & &  \\

%76:63 & not resolved  & & & \\

78:12 & 13/60 & 0.08 & 11 & 16.4\\

89:22 & 8/40 & 0.08 & 11 & 16.3\\

89:33 & 7/11 & 0.05 & 7 &  \\

90:12 & 25/125 & 0.2 & 27 & 19.0\\

%91:41 & 11/41 & 0.08 & 11 & 20.5\\ 

92:74 & not resolved  &  & \\

91:93 & 35/298 & 0.10 & 16 & 15.9\\

FCOS 1-063& 35/91 & 0.08 & 11 & 16.3 \\

92:39 & 2.6/2.4 & stellar & & \\

star & 2.6/2.4 & stellar && \\
\hline
\end{tabular}
\end{table*}

\begin{figure*}
\includegraphics[width=16cm]{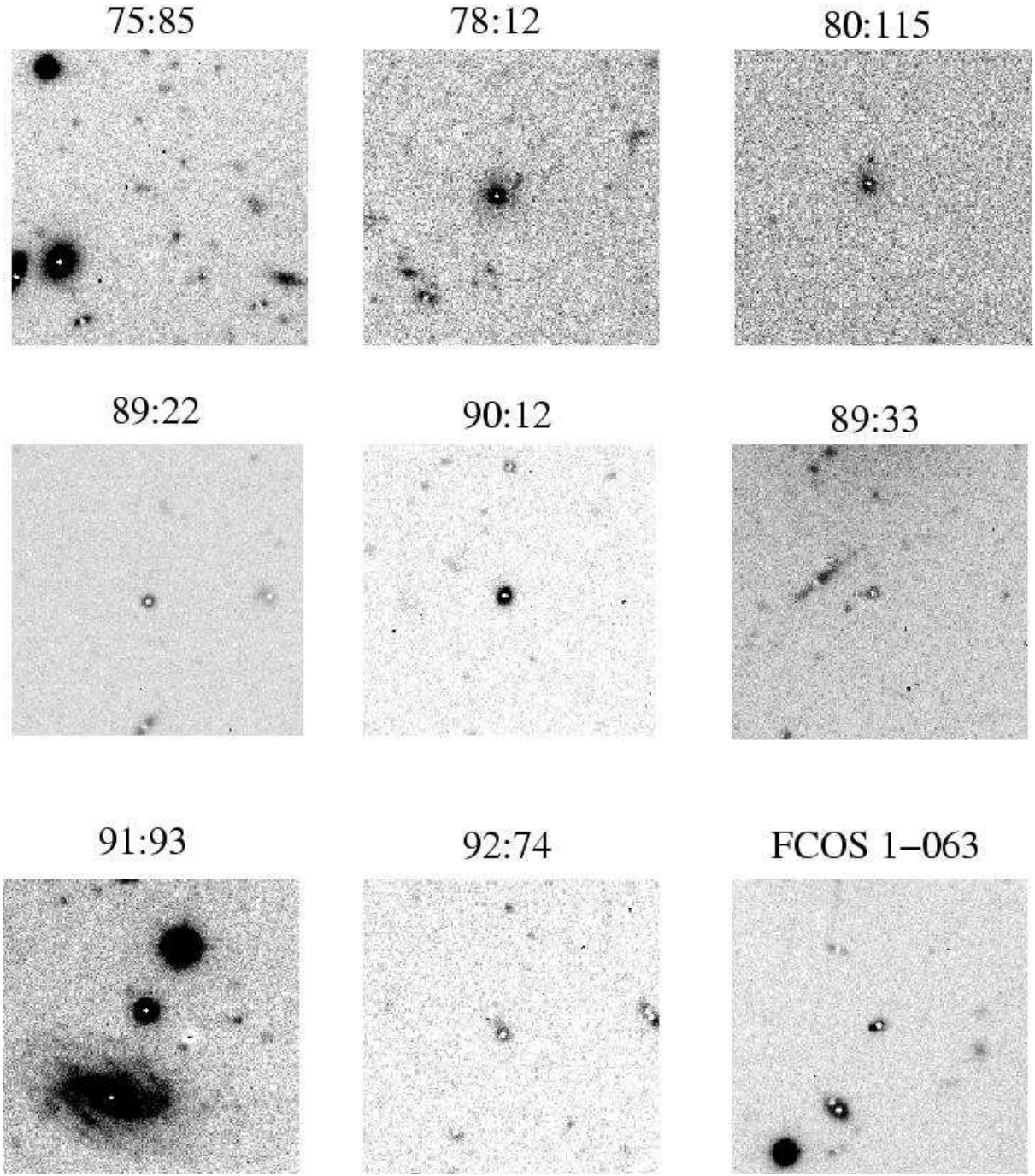}
\caption{Centered on the objects, these images show the residuals of 9 objects after PSF subtraction. The size
of each image is 35$\times$35$\arcsec$. North is up, East to the left. See the text for
individual
descriptions.
}
\label{fig:chartsub}
\end{figure*}

\begin{figure}
\includegraphics[width=9cm,angle=-90]{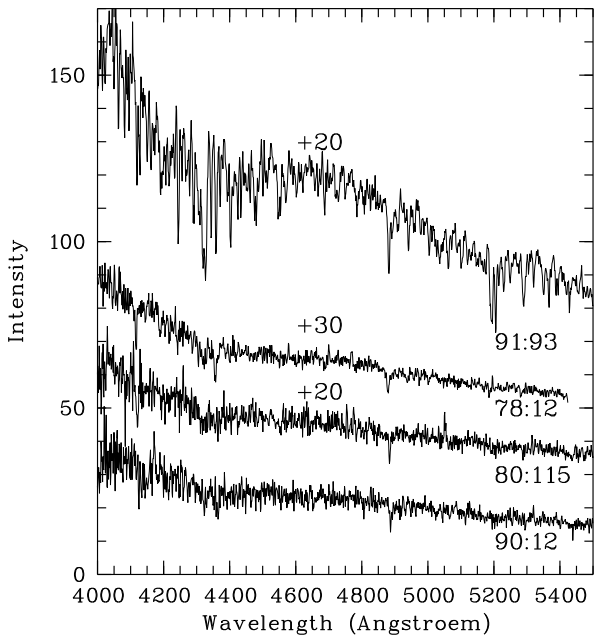}
\caption{Four spectra of the brightest objects are plotted in the wavelength range 4000 \AA - 5500 \AA.
The spectra are not flux calibrated and are shifted (except for 90:12) to permit a convenient display. Individual
shifts are indicated. Except perhaps
for 91:93, they are too noisy to measure line indices. However, the high metallicity of 91:93 and 
the strong Balmer lines of 78:12 are striking. The ''emission'' in 80:115 at 5050 \AA\ is an artefact.
}
\label{fig:spectra}
\end{figure}

\subsection{Description of Objects}
Our spectra were designed to give radial velocities, not line indices. Therefore, most of them are too
noisy to say anything about abundances or ages. In Fig.\ref{fig:spectra} 4 spectra of the brighter
objects are displayed. They are not flux-calibrated (the relative intensity rather reflects the spectrograph's
flat field) and are shifted conveniently. Indicated are the individual shifts. In the following
description, we refer a few times to Balmer line strengths. 

\paragraph{Object 75:85}
This object is marginally resolved by ISHAPE, which may be caused by its 
superposition on a faint, elongated structure, pointing towards N-E. The PSF-subtraction does not show a
 remnant. Washington colours indicate a metallicity of -0.3 dex. 
 Aperture photometry gives for the appendage a V-I colour being 0.2$\pm$0.2 mag
redder than the object itself, still consistent with having the same colour.
However, the appendage is very faint and the
 colour measurement might be not reliable.
Although one cannot exclude that the object is superimposed on a faint
 background galaxy, it is a candidate for a bright globular cluster with a striking tidal tail.

%\subsubsection{76:63}
%We show this object for curiosity. Given the location near to the galaxy
%group, it is very probable that it is superimposed on a background galaxy.
%The metallicity is solar. Aperture photometry indicates a bluer colour of the
%galaxy by 0.3$\pm$0.2 mag. The PSF subtraction leaves no remnant.  

\paragraph{Object 78:12}
This object is stunning. 
The PSF-subtraction uncovers a large halo, extending on our image to about 3$\arcsec$
to the North, corresponding to 270pc, which clearly is a lower limit. 
The effective radius of 11 pc 
only refers to the core. For comparison, $\omega$-Centauri
has an effective radius of 6.4 pc, but with an absolute magnitude of $\mathrm M_V = -10.5$ is only 0.4 mag fainter. The central surface brightness consequently is low, but still normal for globular clusters.  
The V-image shows two faint structures, one going in the N-W direction, the other, like a tiny "spiral arm" to the
 S-W. The N-W structure is also
visible on our MOSAIC R-image. It is hardly visible on the VLT I-image probably due
to the much higher sky background. 
The metallicity
from Washington photometry is -1.3 dex. The spectrum shows strong Balmer lines. H$\gamma$ is much
stronger than the neighbouring G-band, which one rather finds for very metal-poor clusters. However, higher
spectroscopic S/N is  needed for any attempt to determine its age. 
In total, 78:12 does not resemble neither a globular cluster nor a dwarf galaxy and
apparently is a candidate for an initially larger but strongly distorted/stripped nucleated
dwarf galaxy. 

\paragraph{Object 80:115}

This object only appears on our V-preimage, which has been exposed for 30 s (seeing 0.7$\arcsec$). Deeper imaging is not available,
thus the S/N is not sufficient for ISHAPE to work reliably. The PSF subtraction however shows a remnant with
an approximate diameter of 2$\arcsec$. The real extent is presumably much
 larger and may even be larger than that of 78:12 given the short exposure time. 
The Washington metallicity is -1.2 dex and the Balmer lines are less strong than in the case of
78:12. 

\paragraph{Object 89:22}
Washington photometry of this resolved object is not available. The V-I colour
indicates a metal-poor object, while the Balmer lines are only moderately
strong. It looks compact, without a visible halo. The diameter is roughly 
2$\arcsec$, corresponding to 180 pc. The effective radius according to ISHAPE is
as large as that of 78:12. It could be a fainter version of an UCD (Drinkwater et al. \cite{drink2003}). 

\paragraph{Object 89:33}
89:33 is only marginally resolved, as are many objects on our frames. It is interesting
by showing a faint appendage, perhaps a tidal tail? 

\paragraph{Object 90:12}

This is the object for which ISHAPE found the largest FWHM. It appears
slightly elliptical. We estimate its total extension along the major axis
to be at least 350 pc according to the remnant of the PSF subtraction. The effective radius is 27 pc, larger than those of
the majority of the UCDs (Drinkwater et al. \cite{drink2003}). 
Mieske et al. (\cite{mieske2004}) quote 1331$\pm$149 km/s for its radial velocity, deviating by about 2-sigma
 from our value.
 
It
can be seen in Fig.\ref{ishape} that the spherically symmetric model leaves a residual.
 Apparently, the object is asymmetric.
No Washington photometry exists, but among our objects it has the bluest
V-I colour, pointing  to a strong metal deficiency. The spectrum is noisy, but the Balmer lines  
appear not much weaker as those of 78:12.
The spatial resolution is good enough for an estimation of the central surface brightness.  
A value of $\rm \mu_0(V) \approx 19 mag/arcsec^2$ is quite low for a globular cluster. 
We estimate the central mass density using formula (7) of Larsen (\cite{larsen01}) and get 
$\rm \rho_0 \approx 170\,M_{\odot}pc^{-3}$ for an adopted $\rm (M/L)_V$ of 4. 
There are just few galactic 
globular clusters having lower central mass densities. 
This combination
of size and  surface brightness, however, does not exist for any known cluster (see Fig.7 of
Huxor et al. \cite{huxor2004}).

\paragraph{Object 91:93}
This is one of the UCDs (UCD 2), which has
been imaged by HST (Drinkwater et al. \cite{drink2003}, de Propris et al. \cite{depropris2005}). The Washington metallicity is -0.6, which
 excellently fits to the metallicity given by Mieske et al. (\cite{mieske2002}) derived from line indices. 
We show the spectrum for illustration, which is typical for a metal-rich globular cluster.
This object gives us the possibility to compare the effective radius of ISHAPE with the more reliable
one of HST. de Propris et al. (\cite{depropris2005})  quote an effective radius of 20.3 pc
so our value
of 16 pc (which is model dependent) shows that a King profile with c=30 measures
its extension only approximately. 
Fig.\ref{ishape} shows a residual. It is symmetric but our King30 model underestimates the brightness of the central parts
and the outer regions, while overestimating the brightness in an intermediate region.
However, our central surface brightness of $\rm \mu_0(V) \approx 16.3\, mag/arcsec^2$ fits 
fits well to the profile of de Propris et al.

The remnant of the PSF has a diameter of about 3$\arcsec$, corresponding to
280 pc. No halo is visible but there is a faint structure at the N-E, too faint
to measure its colour reliably. Deeper
images could provide colour measurements in order to see whether this is
a faint background galaxy.
Mieske et al. (\cite{mieske2004}) quote 1280$\pm$58 km/s for its radial velocity which agrees with
our value within the uncertainty.

\paragraph{Object 92:74}
The Washington metallicity of this object is $\rm -0.9$. ISHAPE resolves it
 marginally. It is superimposed on a faint elongated structure pointing towards N-E, the
length being approximately 3 arsec. This structure is marginally visible on
our R-frame and not visible on the I-frame.
Like 75:85, it may be a candidate for showing a bright tidal tail. Mieske et al. (\cite{mieske2004}) quote 997$\pm$152 km/s 
for its radial velocity and thus agrees with our more accurate value.

\paragraph{FCOS 1-063}

Mieske et al. (\cite{mieske2002}) list this object as a globular cluster. Our image immediately 
reveals its elongation.
 The subtraction of the PSF further shows that it actually consists of two sources,
superimposed on a larger structure with major and minor
axes of approximately 2.5 and 1.5 arcsec of length, 
corresponding  to 140 kpc and 230 kpc, respectively. 
The secondary object is 2 mag fainter than the primary. No difference
in V-I color can be detected. It still could be a superimposed star but in conjunction
with its other peculiarities it would be a quite strange coincidence. 

The Washington color indicates a metallicity of -1.1. The absolute
magnitude is $M_V = -11$. 
The radial velocity relative to NGC 1399 is almost 800 km/s. If we 
assume (very probable incorrectly) that the radial component represented the full 
orbital velocity  around NGC 1399 the object was at its perigalactic distance. This is the  miminal
space velocity which is possible relative to NGC 1399. Already then  it must have an
 extremely
elongated orbit (Richtler et al. \cite{richtler2004}). The radial separation from NGC 1399 is 6 arcmin, 
corresponding to a perigalactic distance of 33 kpc. The apogalactic distance is then 
about 200 kpc (see Fig.20 of Richtler et al. \cite{richtler2004}).  The true space velocity relative to NGC 1399
probably is higher and the apogalactic distance larger than 200 kpc.
Then the question arises whether such object can be bound to NGC 1399 or rather should be
considered as an interlooper bound only to the entire Fornax cluster.

\section{Discussion}
The main aim of our presentation is  
to call the community's attention to the present objects. Given that almost all have been
found on two FORS fields only after a modest exposure time, the cluster system of NGC 1399 must host a multitud of similar
objects. Extended speculations on the nature of these sources are inappropriate but a few
remarks may be given.
 
 What is the nature of the secondary point source in the case of FCO 1-063? Its
association with the main source is unclear, although likely. It can be a
 candidate for a nucleus
of a stripped
dwarf galaxy with one globular cluster, which survived tidal stripping. 
One is tempted to think of a cluster merging with the nucleus, as predicted by
Oh \& Lin (\cite{oh2000}) and  Lotz et al. (\cite{lotz2001}).
 Long-slit spectroscopy in excellent seeing would prove/disprove its association with FCO 1-063. 

%The mere fact that objects like 78:12 or 80:115 do not resemble any known dwarf galaxy or globular cluster
%already suggests that they  underwent some sort of transformation.  

The interpretation of UCD's discovered by Drinkwater et al. 
(\cite{drink2000}) and Hilker et al. (\cite{hilker1999}) as remnants of
dwarf galaxies may also hold for fainter objects. Depending on the initial configuration and
the details of the interaction process, ''galaxy threshing'' (Karick et al. \cite{karick2003},
Bekki et al. \cite{bekki2001}, \cite{bekki2003})
 may produce  
a large variety of morphological appearance, the UCDs only being the brightest ones.
However, de Propris et al. \cite{depropris2005} compared surface brightness profiles of UCDs and
dwarf galaxy nuclei and found the latter to have systematically lower central surface brightnesses,weakening the case for galaxy threshing.
Regarding their central surface brightness, our objects, except 90:12, resemble  fainter
versions of UCDs rather than dwarf galaxy nuclei. 90:12 and perhaps 80:115 (where deeper observations are required) seem to be quite different from UCDs.

The role of accretion of dwarf galaxy nuclei might thus be  
important
    in giant ellipticals, particularly for a central giant
elliptical like NGC 1399. This is also interesting in the following context: Bimodal colour
 distributions are common in globular cluster systems. However, although the bimodality in
NGC 1399 is very pronounced it disappears for the bright end of the luminosity function (Dirsch et al, 
\cite{dirsch2003})
indicating that this population has a different history than the bulk of the fainter
clusters. The sample of bright compact objects of Mieske et al. (\cite{mieske2004}) emphasizes
this point again.  
 
Another possibility 
is that Blue Compact Dwarf galaxies (BCDs) may be progenitors. An interesting object
in that respect is POX 186 (e.g. Kunth 1981, Corbin \& Vacca 2002, Guseva et al. 2004). Such extremely
compact star-forming dwarf galaxies may well develop into objects like 78:12 or 80:115 after star formation has ceased.
The formation of ''super star clusters'' and perhaps their subsequent merging (Fellhauer \& Kroupa \cite{fellhauer2002}) 
does not need a merger of large spiral galaxies, but also occurs in BCDs (e.g. Vanzi \cite{vanzi2003}).
Dwarf galaxies may therefore contribute in various ways to the population of bright and compact objects.
Also the ''faint fuzzies'' in lenticular galaxies (Brodie \& Larsen \cite{brodie2002}) and the recently discovered
globular clusters with large (30 pc) effective radii in M31 (Huxor et al. \cite{huxor2004}) show that the morphology
of globular clusters is much broader than previously thought. The brightnesses of our objects
are compatible with the bright tail of the luminosity function of GCs in NGC 1399 (e.g. Dirsch 
et al. \cite{dirsch2003}). Once a statistically significant sample of these objects has been
assembled, one can compare their luminosity function with that of GCs (for a recent review
see Richtler \cite{richtler2003}) and search for differences.

It is therefore of high interest to perform a
complete census around NGC 1399. HST with the Advanced Survey Camera would of course be preferable,
but ground-based imaging with a seeing of 0.5" is also feasible and probably would uncover many more
interesting and surprising objects.

\begin{acknowledgements}
We thank an anonymous referee for constructive remarks.
T.R., B.D., and L.I. acknowledge support from the FONDAP center for astrophysics, Conicyt 15010003.

\end{acknowledgements}

%\begin{quote}
{}
%\end{quote}


\begin{thebibliography}{}
\bibitem[1992]{ashman92} Ashman, K. M., Zepf, S.E. 1992, ApJ 384, 50

\bibitem[2001]{bekki2001} Bekki, K., Couch, W.J., Drinkwater, M.J. 2001, ApJ 552, L105

\bibitem[2003]{bekki2003} Bekki, K., Couch, W.J., Drinkwater, M.J., Shioya, J., 2003, MNRAS 344, 399

\bibitem[2003]{brodie2002} Brodie, J.P., Larsen, S.S. 2002, AJ 124, 1410

\bibitem[2002]{corbin2002} Corbin, M. R., Vacca, W. D. 2002, ApJ 581, 1039

\bibitem[1998]{cote1998} C\^ot\'e, P., Marzke, R.O. West, M.J. 1998
 ApJ 501, 554

\bibitem[2005]{depropris2005} de Propris, R., Phillipps, S., Drinkwater, M. et al. 2005, ApJ 623, L105

\bibitem[2003]{dirsch2003} Dirsch, B., Richtler, T., Geisler, D. et al. 2003,
AJ 125, 1908

\bibitem[2004]{dirsch2004} Dirsch, B., Richtler, T., Geisler, D. et al. 2004,
AJ 127, 2114 

\bibitem[2000]{drink2000} Drinkwater, M. J., Jones J.B., Gregg M.D., Phillips S.
2000, PASA 17, 227

\bibitem[2003]{drink2003} Drinkwater, M. J., Gregg, M.D., Hilker, M., et al. 2003,
 Nature 423, 519

\bibitem[2002]{fellhauer2002} Fellhauer, M., Kroupa, P. 2002, MNRAS 330, 642

\bibitem[1997]{forbes97} Forbes, D.A., Brodie, J.P., Grillmair, C.J. 1997, AJ 113, 1652

\bibitem[2004]{guseva2004} Guseva, N.G., Papaderos, P., Izotov, Y.I., Noeske, K.G., Fricke K.J. 2004,
A\&A 421, 519
 

\bibitem[2002]{harris2002} Harris, W.E., Harris, G.L.H. 2002, AJ 124, 1435

\bibitem[2004]{hilker2004} Hilker, M., Kayser, A., Richtler, T., Willemsen P. 2004.
A\&A 422, L9 

\bibitem[1999]{hilker1999} Hilker M., Infante L., Vieira G., et al. 1999, A\&AS  134, 75

\bibitem[1999]{hilker1999a} Hilker, M., Infante, L., Richtler, T. 1999, A\&AS  138, 55

\bibitem[2004]{huxor2004} Huxor, A.P., Tanvir, N.R., Irwin, M.J. et al., MNRAS, in press (astro-ph0412223)

\bibitem[1997]{ibata1997} Ibata, R. A., Wyse, R. F. G., Gilmore, G., et al. 1997,
AJ 113, 634

\bibitem[2003]{karick2003} Karick, A., Drinkwater, M.J., Gregg, M.D. 2003, MNRAS 344,
188

\bibitem[1999]{kissler1999} Kissler-Patig, M., Grillmair, C.J., Meylan, G. et al.
1999, AJ 117, 1206

\bibitem[1981]{kunth1981} Kunth, D., Sargent, W. L. W.,  Kowal, C. 1981, A\&AS 44, 229 

\bibitem[1999]{larsen1999} Larsen, S.S. 1999, A\&A 354, 836

\bibitem[2001]{larsen01} Larsen, S.S. 2001, AJ 122, 1782 

\bibitem[2001]{larsen2001} Larsen, S.S., Brodie J.P., Huchra J.P, et al. 2001, 
AJ 121, 2974

\bibitem[2001]{lotz2001} Lotz, J.M, Telford, R., Ferguson, H., et al. 2001, ApJ 552, 572

\bibitem[2004]{martini2004} Martini, P., Ho, L.C. 2004, ApJ 610, 233

\bibitem[2002]{mieske2002} Mieske, S., Hilker, M., Infante L. 2002, A\&A 383, 823

\bibitem[2004]{mieske2004} Mieske, S., Hilker, M., Infante, L. 2004, A\&A 418, 445

\bibitem[2000]{oh2000} Oh, K.S., Lin, D.N.C. 2000, ApJ 543, 620

\bibitem[2001]{phillips01} Phillips, S, Drinkwater, M.J., Gregg, M.D. \& Jones J.B. 2001,
ApJ 560, 201

\bibitem[2003]{richtler2003} Richtler, T. 2003, in "Stellar Candles for the Extragalactic Distance Scale", Eds. D. Alloin and W. Gieren, Springer, Lecture Notes in Physics, vol. 635, p.281\\

\bibitem[2004]{richtler2004} Richtler, T., Dirsch, B., Gebhardt K.,  et al. 2004,
AJ 127, 2094

\bibitem[2003]{vanzi2003} Vanzi, L. 2003, A\&A 408, 523

\bibitem[1988]{zinnecker1988} Zinnecker, H., Keable, C.J., Dunlup, J.S. et al. 1988,
in IAU Symp. 126, Globular Cluster Systems in Galaxies, eds. J. E. Grindlay \& A. G. D. Philip,
p. 603
\end{thebibliography}
\end{document}